\documentclass[
reprint,
twocolumn,
amsmath,amssymb,superscriptaddress,aps,
prl]{revtex4-1}

\usepackage{graphicx}
\usepackage{dcolumn}
\usepackage{bm}

\usepackage[utf8]{inputenc}
\usepackage[T1]{fontenc}
\usepackage{mathptmx}
\usepackage{etoolbox}
\usepackage{xcolor}

\newcommand{\evdel}[1]{\langle #1 \rangle}
\makeatletter
\def\@email#1#2{%
 \endgroup
 \patchcmd{\titleblock@produce}
  {\frontmatter@RRAPformat}
  {\frontmatter@RRAPformat{\produce@RRAP{*#1\href{mailto:#2}{#2}}}\frontmatter@RRAPformat}
  {}{}
}%
\makeatother

\begin{document}

\preprint{AIP/123-QED}

\title[Emergence and co-existence of periodic and unstructured motion in future-avoiding random walks]
{Emergence and co-existence of periodic and unstructured motion in future-avoiding random walks}

\author{A. Schmaus}
 \affiliation{Potsdam Institute for Climate Impact Research,Telegrafenberg A 31, Potsdam, Germany}
 \affiliation{Technical University Berlin, Straße des 17. Juni 135, Berlin 10623, Germany}
 
\author{K. Stiller}%
 
 \affiliation{Potsdam Institute for Climate Impact Research,Telegrafenberg A 31, Potsdam, Germany}
\affiliation{Technical University Berlin, Straße des 17. Juni 135, Berlin 10623, Germany}

\author{N. Molkenthin}
\email{nora.molkenthin@pik-potsdam.de}
 \affiliation{Potsdam Institute for Climate Impact Research,Telegrafenberg A 31, Potsdam, Germany}

\date{\today}

\begin{abstract}
Self-avoiding random walks on graphs can be seen as walkers interacting with their own past history. Here we consider a complementary class of dynamics: Mutual future avoiding random walks (MFARWs), where stochastically driven walkers are avoiding each others planned future trajectories. Such systems arise naturally in conceptual models of shared mobility. We show that periodic behavior emerges spontaneously in two different MFARWs, and that periodic and unstructured behavior coexist, providing a first example of Chimera like behavior of non-oscillatory paths on networks. Further, we analytically describe and predict the onset of structure. We find that the transition from unstructured to periodic behavior is driven by a novel mechanism of self-amplifying coupling to the periodic components of the stochastic drivers of the system. In the context of shared mobility applications, these states imply a regime of naturally stable co-existence between flexible and line-based public transport.
\end{abstract}

\maketitle

\textit{Introduction--} 
Interacting trajectories on networks have been explored both mathematically (e.g.self-avoiding random walks \cite{amit1983asymptotic}) and as models for physical processes (e.g. polymer chains \cite{bhattacharjee2013flory, molkenthin2016scaling} or Levy flight foraging \cite{james2011assessing}). In light of the global challenge of climate change and the rise of the "sharing economy", on-demand ride pooling systems \cite{creutzig2024shared} have gathered the attention of the research community. They too constitute a class of interacting trajectories on networks. While mostly studied for their practical relevance, i.e. finding the best algorithm for pooling \cite{alonso2017demand,horn2002fleet} or analyzing conditions for customer acceptance \cite{shi2021influence,carrion2012value}, some studies have analyzed ride pooling as a complex system and found interesting scaling dynamics \cite{muhle2023analytical, tachet2017scaling, molkenthin2020scaling, zech2022collective} and network structures \cite{bujak2023network}. 

If a system of identical units governed by symmetric rules spontaneously splits into coexisting coherent and non-coherent subpopulations, this is called a chimera state. While introduced in the analysis of coupled oscillators \cite{abrams2004chimera}, the concept has significantly broadened since \cite{haugland2021changing}. Chimeras have been identified, among others, in neural models \cite{hizanidis2016chimera}, fluid flows \cite{nicolaou2017chimera}, and different wave patterns in reaction-diffusion systems \cite{halatek2018rethinking,PhysRevLett.133.207203}, raising fundamental questions about symmetry breaking, multi-stability, and collective behavior in high-dimensional dynamical systems. In all of the examples above, the oscillatory dynamics are characteristics of the media or individual units. In the ride-pooling system, the oscillatory dynamics arise from future-avoiding coupling and its interaction with the network structure.

In the context of mobility, coherent or structured routes equate to line services (i.e. public transport), periodically visiting the same sequence of nodes. Conversely, non-coherent or unstructured routes correspond to flexible trajectories without periodicity, as commonly found in on-demand systems. The interdependence of structured and unstructured routes in human mobility is subject to an ongoing debate. This includes the question whether flexible pooling services are beneficial in reducing the negative impacts of traffic \cite{diao2021impacts,henao2019impact,SCHMAUS2025104726} and how to encourage effective pooling \cite{fielbaum2022split,de2023ride}.

\begin{figure}[b]
    \centering
    \includegraphics[width=\columnwidth]{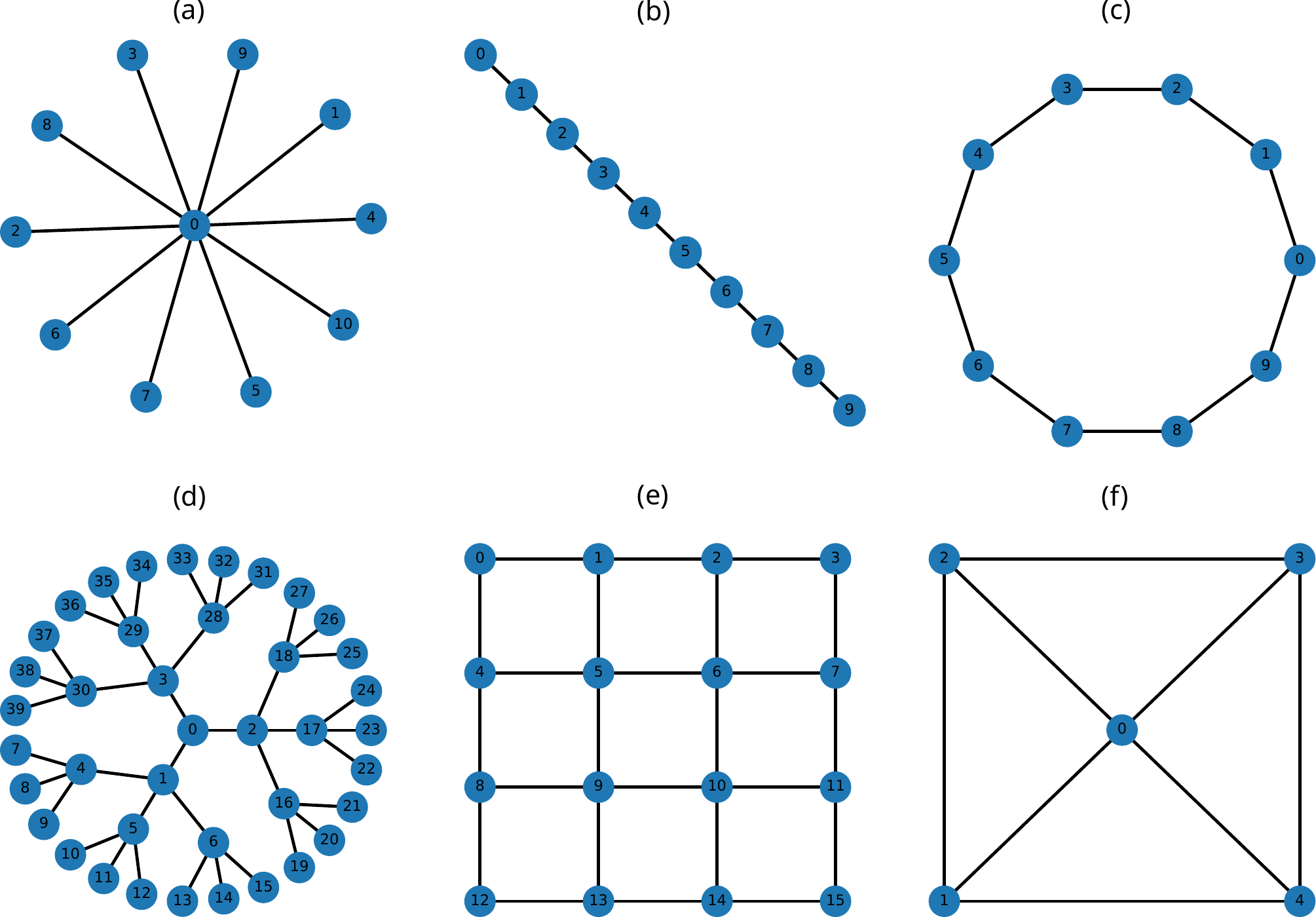}
    \caption{Networks used in this paper: a) star graph with 10 nodes, b) line graph with 10 nodes, c) ring graph with 10 nodes, d) Cayley tree with 40 nodes, e) grid graph with 16 nodes, f) Wheel graph with 5 nodes. Driving from one node to a neighbouring node takes one time unit.}
    \label{fig:networks}
\end{figure}

Here, we analyze simulated fleets of future-avoiding walkers, which are inspired by ride-pooling fleets, for spontaneous emergence of periodic routes. 

We find evidence for the emergence of chimera-like coexistence of coherent and non-coherent states in numerical simulations, and derive an analytic approximation that describes the onset of structure in the paths. The results imply a self-reinforcing mechanism, which preferentially adds detours to shorter routes, leaving the straight additions for longer, straighter routes.

\textit{Methods--} The routes presented in this study are generated from simulations with the python package \textit{ridepy} \cite{Jung2024}. As input the simulation requires a stop network together with timed origin-destination pairs. The stop networks used are shown in Fig.~\ref{fig:networks}. The origin-destination (OD) pairs are uniformly random pairs of nodes. The occurrence times of requests are generated using a Poisson process, with rate $\lambda=\Delta t ^{-1}$, where $\Delta t$ is the average time between requests. To facilitate the comparison between different parameter settings, we use the normalized request load $x$ \cite{molkenthin2020scaling} as: 
\begin{equation}
x= \frac{\lambda \langle l \rangle}{B},
\end{equation}
where $B$ denotes the number of routes in the system and $\langle l \rangle$ the average shortest path length in the network. Routes are generated starting from an initial location by successively including new requests in the route. Where and how a new request is added depends on the \textit{unique stop lists} 
\begin{equation}
    U_b(\tau)=\{u_0,u_1,u_2,...,u_i\}
\end{equation}
at the time of request placement $\tau$, for each route $b \in \{0,1,...,B\}$, which are lists of consecutive stops, without a consecutive repetition of the same node if multiple pick-ups or drop-offs are carried out at the same stop. Note that the unique stop list describes only the future part of the route that has \underline{not yet} been served.

The route $T_b=\{u_{initial},...,u_{final}\}$ denotes the entire trajectory of unit $b$ over the course of the whole simulation. $|T_b|$ denotes the length of the entire trajectory. 

The simulations process the requests sequentially using two dispatcher algorithms: the No-Detour-Heuristic dispatcher (NDH) and the Minimal-Fleet-Distance dispatcher (MFD). The NDH dispatcher is based on the dispatcher used in \cite{manik2020topology}. It identifies the route and the insertion index pair (i.e. pair of indices $(i_o,i_d)$ in stop list), which serves the request with the earliest delivery time under the constraints that the travel time of passengers already assigned is never increased and the vehicle occupancy does not exceed the vehicle capacity. See the Supplementary Material \cite{supplement} for more details. The MFD dispatcher from the \textit{ridepy} package \cite{Jung2024} performs a brute force evaluation of all insertions $(i_o,i_d)$, which fulfill the constraints set for maximum waiting time $t^{\text{max}}_{w}$, maximum delay factor $d=t^{\text{max}}_{d}/t^{\text{direct}}$ or maximum delay $t^{\text{max}}_{d}$ and vehicle capacity $c$ \cite{SCHMAUS2025104726}. Instead of request arrival time, here the added distance to the system serves as the cost function. Note that in both cases, the trajectories interact by competing for requests, which is decided on the basis of the future part of the route.

The \emph{link usage} $\rho(i)$ of edge $i$ is defined as the fraction of edge $i$ in the edges that form any route in the fleet $\mathcal{T}=\{T_0,...,T_B\}$.
If two consecutive stops in the unique stop list are not direct neighbours, the occurrence is evenly divided over all shortest paths between them, however, this is increasingly rare as load $x$ increases.

The final trajectories $T_b$ from simulations and theoretical models are also analyzed using recurrence plots \cite{marwan2007recurrence}. 
To identify recurring nodes on a route and find periodic patterns, we go beyond the binary recurrence matrix and replace the ones of the recurrence matrix with the trajectory value corresponding to these indices. This is related to chromatic recurrence plots \cite{cox2016chromatic} or symbolic recurrence plots \cite{caballero2018symbolic}. We call this the \emph{value-weighted recurrence matrix}:
\begin{equation}
    V_{i,j}=
    \begin{cases}
        u_i \text{     if  } u_i=u_j\\
        0 \text{       if  } u_i \neq u_j
    \end{cases}.
    \label{eq.valrec}
\end{equation}
The valued version $V_{i,j}$ directly visualizes the recurring nodes or regions. 
\begin{figure}[t]
    \centering
\includegraphics[width=\columnwidth]{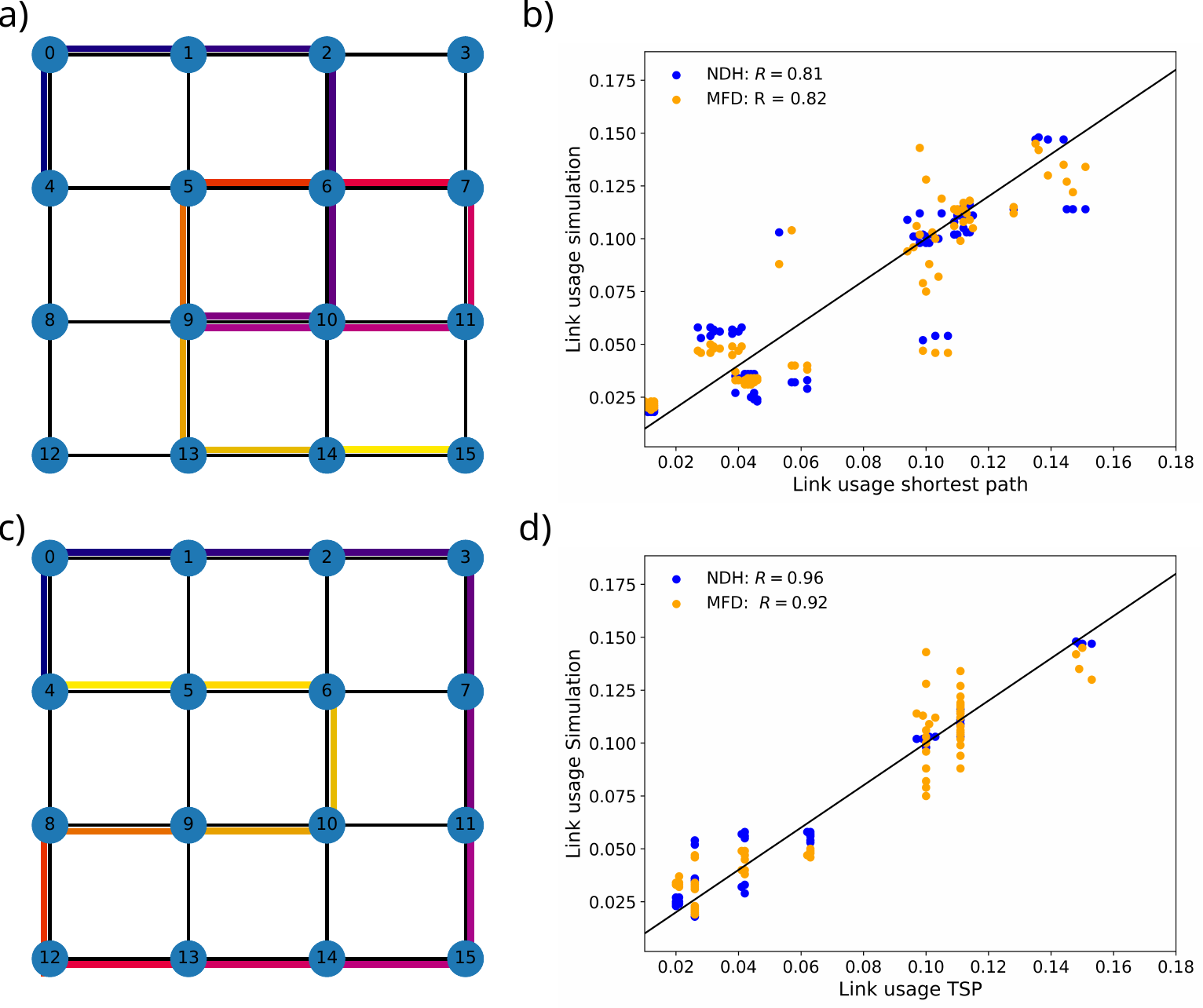}
    \caption{\textbf{TSP tours predict link usage better than random shortest path walks.} a) Example for shortest path walk on 4x4 grid, visiting the nodes (4,2,9,11,6,13,15). b) The edge betweenness or link usage for a perfect taxi service predicts the link usage observed in the simulations with $R^2=0.81$ for the NDH and $R^2=0.82$ for the MFD dispatcher for a single vehicle and a load of $x=100$. Each point refers to one edge in one of the networks from Fig.~\ref{fig:networks}. c) One of the shortest tours visiting each point in the 4x4 grid at least once. d) The mix of all TSP tours including rotations and mirrored versions predicts link usages with $R^2=0.96$ or $R^2=0.92$ respectively.}
    \label{fig:linkusage1}
\end{figure}

To quantify periodicity found in the recurrence plots we use the \emph{average diagonal line length}, as defined in \cite{marwan2007recurrence}:
\begin{equation}
    L=\frac{\sum_{l_{min}}^{N}l P(l)}{\sum_{l=1}^{N}P(l)},
    \label{eq.L}
\end{equation}
where we chose $l_{min}=2$ and $P(l)$ is the histogram of diagonal line lengths.


\textit{Periodic routes explain single route link usage patterns--} Even for a single route, the dynamics show significant structural richness, with traveling salesperson (TS) like paths seemingly playing a central role. At very low loads $x \ll 1$, each route is a sequence of shortest paths along the network connecting uniformly random node pairs Fig.~\ref{fig:linkusage1} a) shows an example. The \emph{shortest path walk} is thus defined as a the shortest path connecting a sequence of uniformly random nodes. Its link usages can be shown to be equal to the normalized edge betweenness centrality (c.f. Supplemental Material).

Fig.~\ref{fig:linkusage1} b) shows the link usages resulting from the simulations using both dispatchers on all six topologies over the edge betweenness centrality of the respective edges for the high load $x=100$. Although it shows a clear correlation between link usage and edge betweenness, with a coefficient of determination of $R^2=0.81$ for the NDH dispatcher and $R^2=0.82$ for the MFD dispatcher, the null model clearly does not fully explain the data.
\begin{figure*}[t]
    \centering
\includegraphics[width=1.8\columnwidth]{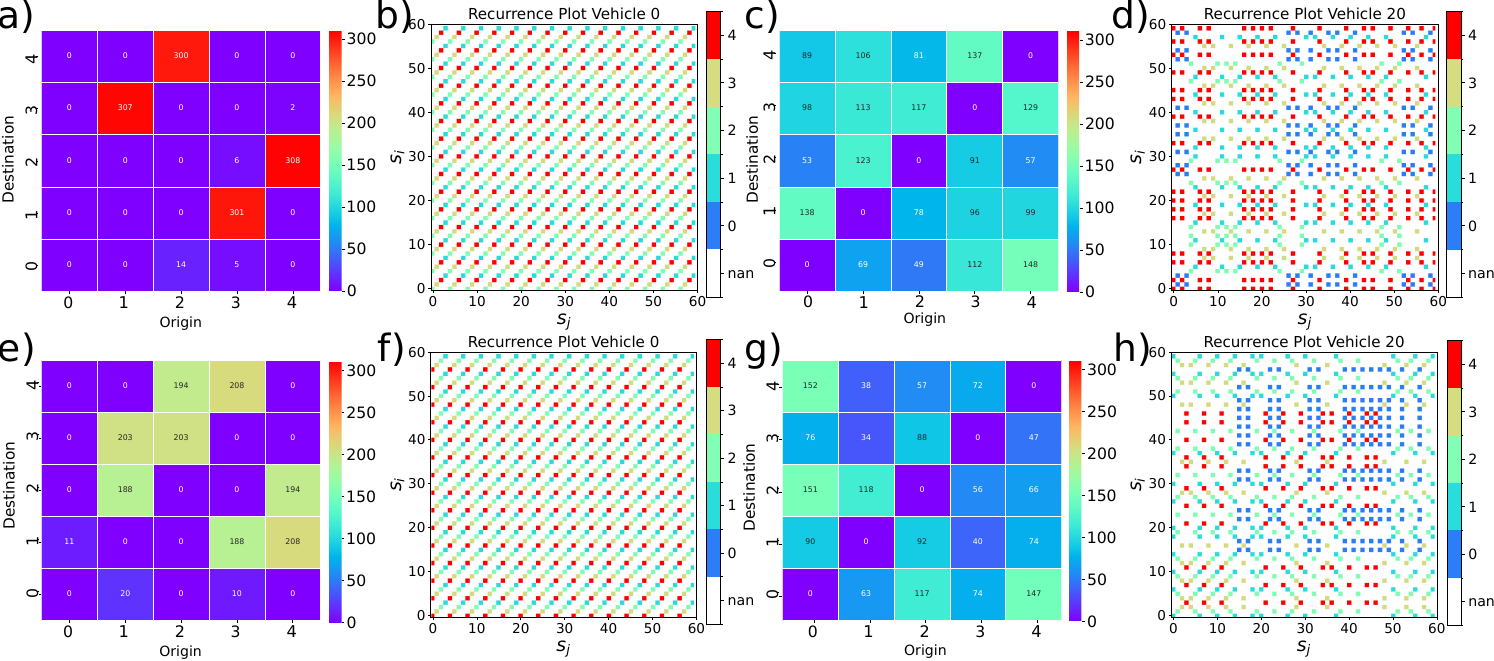}
    \caption{\textbf{Recurrence plots reveal periodic and unstructured routes.} All results are from tests with $B=100$, $x=10$ and a vehicle capacity of 20. a) Accepted request OD-pairs of periodic route (index 0) from MFD simulation. b) Corresponding recurrence plot of route a). c) Accepted request OD-pairs of unstructured route (index 20) from MFD simulation. d) Corresponding recurrence plot of route c). e) Accepted request OD-pairs of periodic route (index 0) from NDH simulation. f) Corresponding recurrence plot of route e). g) Accepted request OD-pairs of unstructured route (index 20) from NDH simulation. h) Corresponding recurrence plot of route g).}
    \label{fig:rec}
\end{figure*}
This implies that at very high loads $x \gg 1$, pooling routes cover the network more systematically. As a model for this, we propose the \emph{stochastic TSP walk} as a route composed of a series of randomly selected segments of traveling salesperson paths of all nodes on the network. Fig.~\ref{fig:linkusage1} c) shows an example.

Fig.~\ref{fig:linkusage1} d) shows the link usages resulting from the simulations over the link usages assuming a stochastic TSP walk of the respective edges. We find that this correlates better with the simulated link usages with a $R^2=0.96$ for the NDH dispatcher and $R^2=0.92$ for the MFD dispatcher. We conclude that the simulated routes of a single trajectory are more structured than the uniformly random origin-destination pairs would suggest and show indication of intermitted periodic behaviour in both algorithms.

\textit{Emergence of chimeras in fleet simulations--} Simulating fleets of $B$ units on network (f) from Fig.~\ref{fig:networks}, creates $B$ routes, which interact indirectly by competing for the assignment of requests. The individual routes all follow identical dynamics with non-linear, homogeneous all-to-all coupling. The resulting routes self-organize into distinctly different movement patterns, as seen in  Fig.~\ref{fig:rec}. The average shortest path length is $\langle l \rangle = 1.2$. With $x=10$ we get $ \frac{1}{\lambda} = \Delta t = 0.00012$. Thus, for 100,000 requests, we simulate a time span of 120 time units, as crossing one edge takes one time unit, this corresponds to route lengths of at most 120 edges.

The routes depicted in Fig.~\ref{fig:rec} a, b, e, and f) are periodic over the entire simulation time. This is evidenced by the dominance of diagonal lines in the recurrence plots in Fig.~\ref{fig:rec} b) and f). In contrast, the routes in Fig.~\ref{fig:rec} c, d and g, f) show no long diagonals in the recurrence plots, indicating unstructured behaviour. Fig.~\ref{fig:rec} a, c, e and g) shows the accepted origin-destination (OD) pairs of one route and reveals different patterns in the allocation of requests to routes. The MFD dispatcher reserves the periodic routes for the "long-distance-requests" going from corner to opposite corner, as shown in Fig.~\ref{fig:rec} a). The NDH dispatcher on the other hand equally selects all OD-pairs on the periodic route to be served by a periodic vehicle. Fig.~\ref{fig:rec} c and g) show that all OD pairs are assigned to the unstructured routes with approximately constant probabilities.

The average diagonal line length $L$, as defined in Eq.~\ref{eq.L} is proportional to the average duration of route periodicity for each trajectory. Since route lengths vary with the underlying network, load and fleet size, we define the normalized average diagonal line length as a measure of \emph{route periodicity fraction}
\begin{equation}
    F=\frac{L}{\evdel{|T|}/2},
\end{equation}
where $\evdel{|T|}$ is the system average over all route lengths and $\evdel{|T|}/2$ is the largest value the average diagonal line length can take for a route of average length with persistent periodic behavior throughout the simulation time.

\begin{figure}[t]
    \centering
\includegraphics[width=\columnwidth]{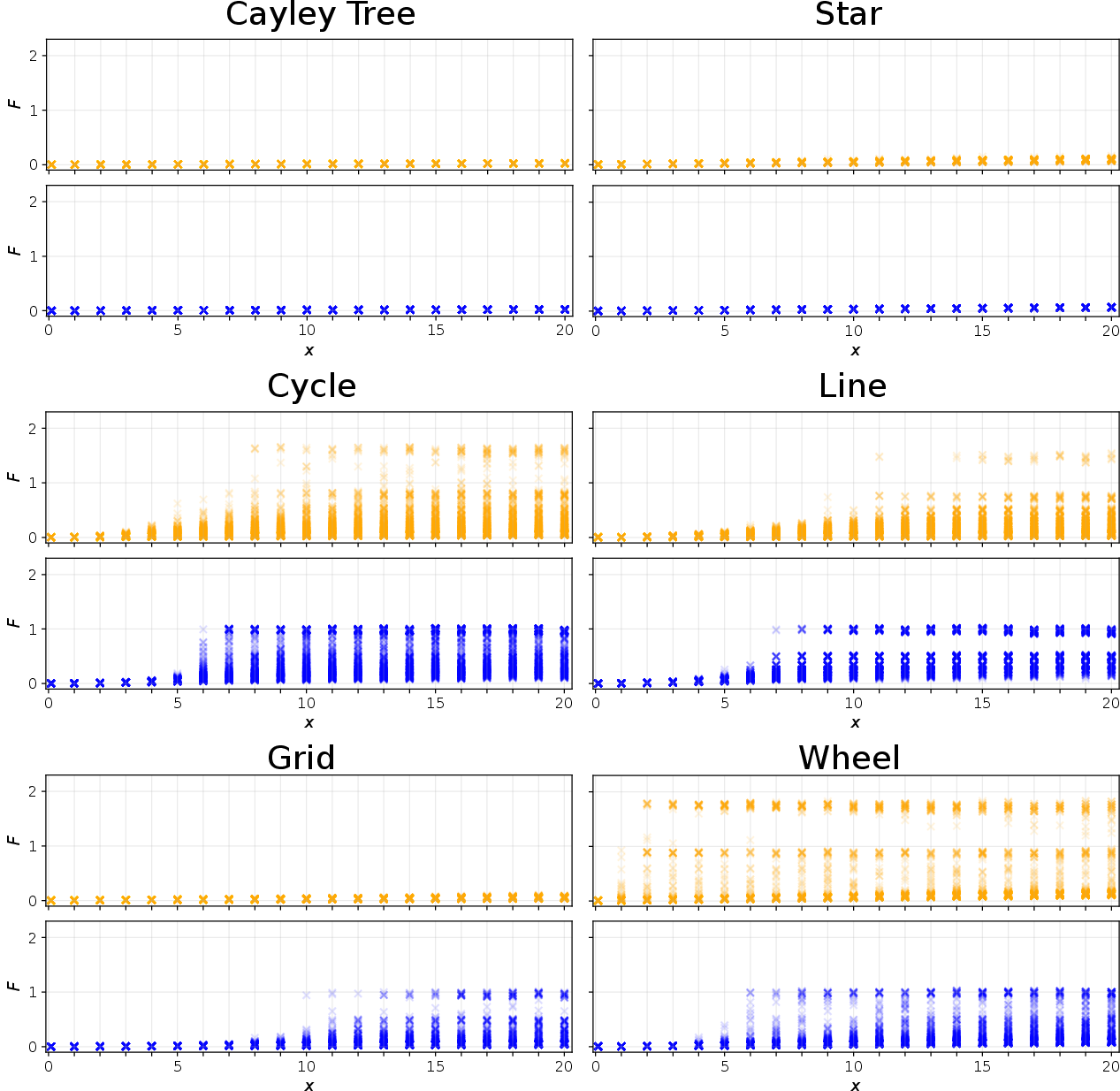}
    \caption{\textbf{Persistent periodic patterns emerge in some cases.} Fraction of route periodicity for each vehicle of the fleet evaluated on all 6 networks for both dispatcher algorithms across loads from $x=0.1$ to $x=20$. Across all simulations $B=100$ and $N_{req}=100000$. Persistent periodic routes emerge for cycle, line and wheel network in both dispatchers and for the grid only in the NDH dispatcher, while routes always remain unstructured in the Cayley tree and star network.}
    \label{fig:chimeras}
\end{figure}

Fig.~\ref{fig:chimeras} visualizes $F$ for both dispatchers and values of the load ranging from $x=0.1$ to $x=20$. For the high-dimensional networks (Cayley tree and star graph, Fig.~\ref{fig:chimeras} top row), $F$ remains low throughout the fleet for all values of $x$ and both dispatchers. In the 1-dimensional networks (cycle and line, Fig.~\ref{fig:chimeras} middle row) $F$ is small for low loads and increases sharply to $F \geq 1$ at intermediate loads. The exact position and sharpness of the increase depends on network, dispatcher, number of simulation steps and fleet size. Interestingly, the route periodicity fraction reaches values above one for the MFD dispatcher (yellow), implying that periodic routes tend to be significantly longer than their unstructured counterparts. This is consistent with the dispatcher's strategy of maximizing idleness to minimize total distance. 
In the intermediate networks (grid and wheel, Fig.~\ref{fig:chimeras} bottom row), we observe mixed behaviours. For the MFD dispatcher on the grid, no globally periodic routes are observed in the given x-range, on the wheel and with the NDH dispatcher on the grid, long-range periodicity emerges.

\begin{figure}[t]
    \centering
\includegraphics[width=\columnwidth]{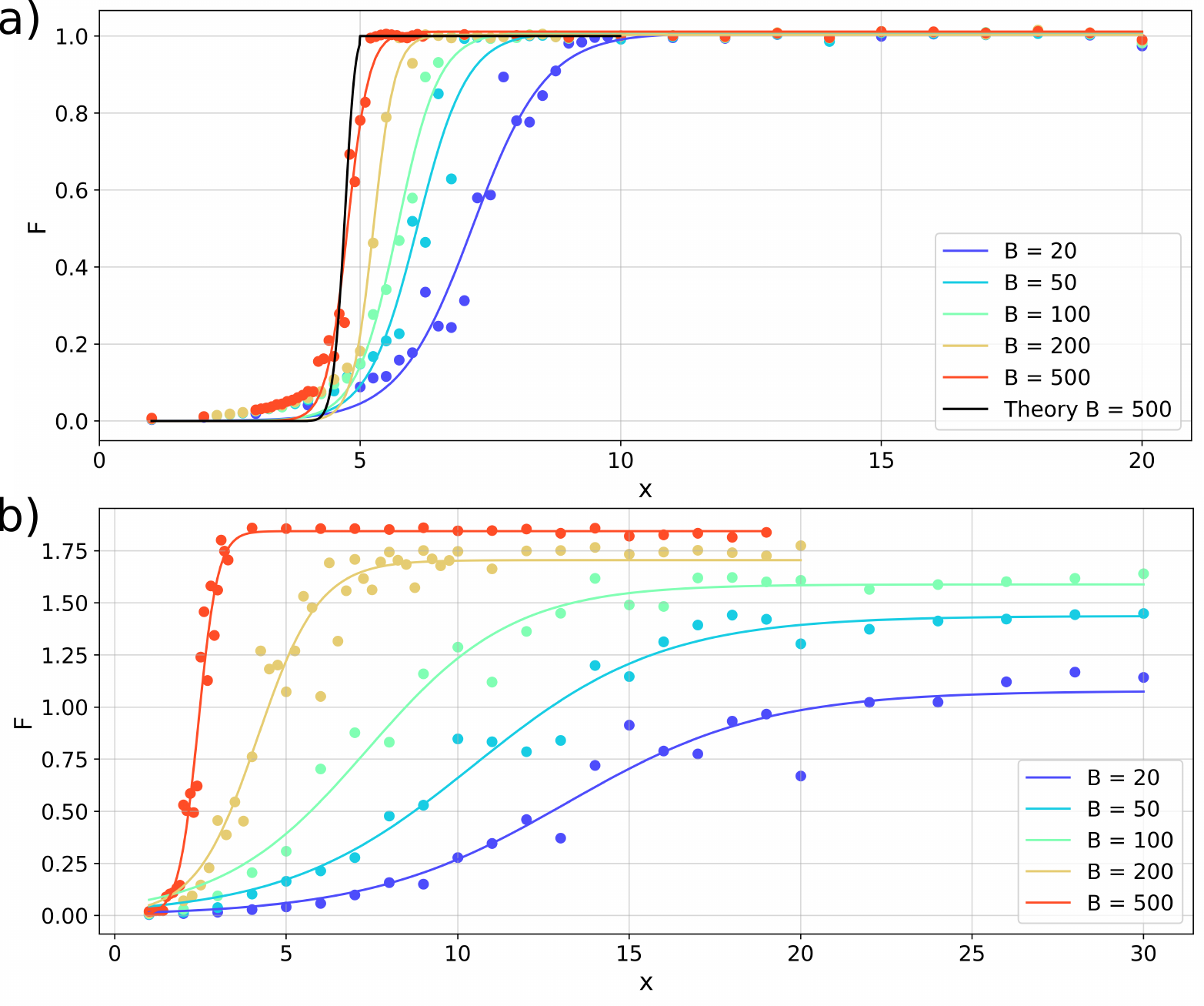}
    \caption{\textbf{Chimera states emerge once periodicity is statistically inevitable.} a) Transition to partially periodic states steepens with fleet size for NDH dispatcher for different fleet sizes $B$ (blue to red dots) and the theoretical approximation for $B=500$ (solid black line). b) Transition to partially periodic states steepens with fleet size for MFD dispatcher.}
    \label{fig:trans}
\end{figure}

\textit{Analytic description of the phase transition on the ring--} To better understand the nature of the phase transitions in Fig.~\ref{fig:chimeras}, we analyze the scaling behaviour of the maximal periodicity fraction $\evdel{F_{\text{max}}(x)}$ averaged over five realizations for fleet sizes of $B=\{20,50,100,200,500\}$ on the cycle graph, with the average number of requests per route kept constant.
The results are shown in Fig.~\ref{fig:trans} along with a mean-field approximation for the NDH dispatcher. 

The approximation is based on the estimation of the probability distribution $P_c$ for a vehicle to change direction when including a new request in its route. This approximation is performed in detail in the Supplemental Material. The main mechanism is to compare the planned length of the nearest route going in the wrong direction along the ring to the waiting time for the nearest route in the correct direction (all other routes are guaranteed to have later arrival times). The longer a currently planned route is, the less likely it is to change direction. This leads to the equation for the direction change probability
\begin{align}
    P_{c}=\int_0^{2M}\prod_{i=1}^{\left \lceil{x}\right \rceil} \frac{s+i v \Delta \tau }{M} \frac{\frac{B}{2}-1}{2M} (1-\frac{s}{2M})^{\frac{B}{2}-2} ds ,
\end{align}
where $M$ is the number of nodes on the ring, $\Delta \tau$ is the average time between requests being assigned to each route.

As a result, some routes accumulate reversals, while others establish periodicity.
This is in line with research on chimeras, such as \cite{zhang2021mechanism}, where the authors argue that the incoherent input from the de-synchronized part of the ensemble stabilizes the synchronous part. 

We use this insight to approximate the periodicity fraction $F$ by identifying it as the probability that a route is periodic over the entire simulation. This probability can be estimated by evaluating the probability of direction changes for $r$ requests for each of the $B$ routes in the system.
\begin{equation}
    F \approx P_{p}=1-(1-(1-P_{c})^{r})^B.
    \label{eq.Fapprox}
\end{equation}

The approximation is plotted in Fig.~\ref{fig:trans} a) for $B=500$ as a black line. Although the location of the transition is accurately predicted and the trend of steepening with fleet size is visible in the simulation results, the steepness of the transition is overestimated as a result of the simplifying assumptions. The functional form of Eq.\ref{eq.Fapprox}, as well as the observed steepening of simulation results suggest a sharpening to a true thermodynamic phase transition in the limit of large fleet sizes.
The results for the MFD dispatcher simulations shown in Fig.~\ref{fig:trans} b) show a similar trend of the transition steepening with increasing $B$.

\textit{Discussion and conclusion--} The above analysis demonstrates that periodic trajectories spontaneously emerge from future-driven interacting path dynamics with uniformly distributed origin-destination pairs for two distinct dispatching dynamics. This is a new type of behaviour that has not been found in other path dynamics (i.e. self-avoiding random walks). Depending on the underlying network and the request density as well as the number of requests and the fleet size, there may be i) only unstructured routes or ii) coexisting periodic and unstructured routes. Although these general categories were observed in both dispatcher dynamics, the conditions under which they occurred varied.

This finding constitutes the discovery of chimera-like states in a novel type of not intrinsically oscillatory dynamics with nonlinear coupling. The analytic approximation for the cycle graph suggests a mechanism of preferential addition of detours to shorter routes, while longer routes attract shortest path additions.

The self-organized emergence of coexisting periodic and unstructured routes is especially interesting, as it hints at the possibility of a natural state of co-existence between flexible and line-based transport options, contributing to the debate whether shared pooled mobility competes with or complements public transport \cite{cats2022beyond,shi2021influence}. In \cite{shi2021influence} the authors highlight that ride-hailing competes with traditional transport modes. The key difference between routes in ride-hailing and ride-pooling lies in the overlapping structure of the latter, increasing route memory and coordination. This results in the observed partially periodic behaviour, which we interpret as synergetic co-existence of fixed and flexible services.

In realistic settings, however, fixed and flexible units often differ in price, vehicle size and comfort, making them harder to compare directly or use interchangeably. Furthermore, most realistic line service systems require vehicle changes.
This raises the question how the emergent structures would change if the dynamics would allow vehicle changes as part of the trip and if this would lead to the emergence of feeder systems such as in \cite{fielbaum2024design}. 

The flexibility of systems with emergent periodicity may offer additional benefits, such as a natural response to system changes and disruptions by flexibly changing the ratio of flexible to periodic units. This kind of flexible dual mode transport systems could be more resilient to disruptions and catastrophes than traditional public transport systems, as well as offering different options to populations with heterogeneous transport preferences \cite{zhong2026beyond}. 

\textit{Data Availability}
The code used to analyze the data is available from \cite{schmaus_2026_20112639}.

\begin{acknowledgments}
\textit{Acknowledgments.} We thank Malte Schroeder and Norbert Marwan for fruitful discussions. Alexander Schmaus acknowledges support from the German Federal Environmental Foundation (Deutsche Bundesstiftung Umwelt).
\end{acknowledgments}

\bibliographystyle{unsrt}
\bibliography{ridepool}

\begin{thebibliography}{10}

\bibitem{supplement}
See supplemental material [url] for an extended description of the dispatchers, analysis at lower loads and an extended derivation of the theoretical approximation.

\bibitem{abrams2004chimera}
D.~M. Abrams and S.~H. Strogatz.
\newblock Chimera states for coupled oscillators.
\newblock {\em Physical review letters}, 93(17):174102, 2004.

\bibitem{alonso2017demand}
J.~Alonso-Mora, S.~Samaranayake, A.~Wallar, E.~Frazzoli, and D.~Rus.
\newblock On-demand high-capacity ride-sharing via dynamic trip-vehicle assignment.
\newblock {\em Proceedings of the National Academy of Sciences}, page 201611675, 2017.

\bibitem{amit1983asymptotic}
D.~J. Amit, G.~Parisi, and L.~Peliti.
\newblock Asymptotic behavior of the" true" self-avoiding walk.
\newblock {\em Physical review B}, 27(3):1635, 1983.

\bibitem{bhattacharjee2013flory}
S.~M. Bhattacharjee, A.~Giacometti, and A.~Maritan.
\newblock Flory theory for polymers.
\newblock {\em Journal of Physics: Condensed Matter}, 25(50):503101, 2013.

\bibitem{bujak2023network}
M.~Bujak and R.~Kucharski.
\newblock Network structures of urban ride-pooling problems and their properties.
\newblock {\em Social Network Analysis and Mining}, 13(1):89, 2023.

\bibitem{caballero2018symbolic}
M.~Caballero-Pintado, M.~Matilla-Garc{\'\i}a, and M.~Ruiz~Mar{\'\i}n.
\newblock Symbolic recurrence plots to analyze dynamical systems.
\newblock {\em Chaos: An Interdisciplinary Journal of Nonlinear Science}, 28(6), 2018.

\bibitem{carrion2012value}
C.~Carrion and D.~Levinson.
\newblock Value of travel time reliability: A review of current evidence.
\newblock {\em Transportation research part A: policy and practice}, 46(4):720--741, 2012.

\bibitem{cats2022beyond}
O.~Cats, R.~Kucharski, S.~R. Danda, and M.~Yap.
\newblock Beyond the dichotomy: How ride-hailing competes with and complements public transport.
\newblock {\em Plos one}, 17(1):e0262496, 2022.

\bibitem{cox2016chromatic}
R.~F. Cox, S.~van~der Steen, M.~Guevara, L.~de~Jonge-Hoekstra, and M.~van Dijk.
\newblock Chromatic and anisotropic cross-recurrence quantification analysis of interpersonal behavior.
\newblock In {\em Recurrence Plots and Their Quantifications: Expanding Horizons: Proceedings of the 6th International Symposium on Recurrence Plots, Grenoble, France, 17-19 June 2015}, pages 209--225. Springer, 2016.

\bibitem{creutzig2024shared}
F.~Creutzig, A.~Schmaus, E.~Ayaragarnchanakul, S.~Becker, G.~Falchetta, J.~Hu, M.~Goletz, A.~Gu{\'e}ret, K.~Nagel, J.~Schild, et~al.
\newblock Shared pooled mobility: expert review from nine disciplines and implications for an emerging transdisciplinary research agenda.
\newblock {\em Environmental Research Letters}, 2024.

\bibitem{de2023ride}
A.~de~Ruijter, O.~Cats, J.~Alonso-Mora, and S.~Hoogendoorn.
\newblock Ride-pooling adoption, efficiency and level of service under alternative demand, behavioural and pricing settings.
\newblock {\em Transportation planning and technology}, 46(4):407--436, 2023.

\bibitem{diao2021impacts}
M.~Diao, H.~Kong, and J.~Zhao.
\newblock Impacts of transportation network companies on urban mobility.
\newblock {\em Nature Sustainability}, 4(6):494--500, 2021.

\bibitem{fielbaum2024design}
A.~Fielbaum and J.~Alonso-Mora.
\newblock Design of mixed fixed-flexible bus public transport networks by tracking the paths of on-demand vehicles.
\newblock {\em Transportation Research Part C: Emerging Technologies}, 168:104580, 2024.

\bibitem{fielbaum2022split}
A.~Fielbaum, R.~Kucharski, O.~Cats, and J.~Alonso-Mora.
\newblock How to split the costs and charge the travellers sharing a ride? aligning system’s optimum with users’ equilibrium.
\newblock {\em European Journal of Operational Research}, 301(3):956--973, 2022.

\bibitem{halatek2018rethinking}
J.~Halatek and E.~Frey.
\newblock Rethinking pattern formation in reaction--diffusion systems.
\newblock {\em Nature Physics}, 14(5):507--514, 2018.

\bibitem{haugland2021changing}
S.~W. Haugland.
\newblock The changing notion of chimera states, a critical review.
\newblock {\em Journal of Physics: Complexity}, 2(3):032001, 2021.

\bibitem{henao2019impact}
A.~Henao and W.~E. Marshall.
\newblock The impact of ride-hailing on vehicle miles traveled.
\newblock {\em Transportation}, 46(6):2173--2194, 2019.

\bibitem{hizanidis2016chimera}
J.~Hizanidis, N.~E. Kouvaris, G.~Zamora-L{\'o}pez, A.~D{\'\i}az-Guilera, and C.~G. Antonopoulos.
\newblock Chimera-like states in modular neural networks.
\newblock {\em Scientific reports}, 6(1):19845, 2016.

\bibitem{horn2002fleet}
M.~E. Horn.
\newblock Fleet scheduling and dispatching for demand-responsive passenger services.
\newblock {\em Transportation Research Part C: Emerging Technologies}, 10(1):35--63, 2002.

\bibitem{james2011assessing}
A.~James, M.~J. Plank, and A.~M. Edwards.
\newblock Assessing l{\'e}vy walks as models of animal foraging.
\newblock {\em Journal of the Royal Society Interface}, 8(62):1233--1247, 2011.

\bibitem{Jung2024}
F.~Jung and D.~Manik.
\newblock Ridepy: A fast and modular framework for simulating ridepooling systems.
\newblock {\em Journal of Open Source Software}, 9(97):6241, 2024.

\bibitem{PhysRevLett.133.207203}
B.-W. Li, J.~Xiao, T.-C. Li, A.~V. Panfilov, and H.~Dierckx.
\newblock Self-organized target wave chimeras in reaction-diffusion media.
\newblock {\em Phys. Rev. Lett.}, 133:207203, Nov 2024.

\bibitem{manik2020topology}
D.~Manik and N.~Molkenthin.
\newblock Topology dependence of on-demand ride-sharing.
\newblock {\em Applied Network Science}, 5(1):1--16, 2020.

\bibitem{marwan2007recurrence}
N.~Marwan, M.~C. Romano, M.~Thiel, and J.~Kurths.
\newblock Recurrence plots for the analysis of complex systems.
\newblock {\em Physics reports}, 438(5-6):237--329, 2007.

\bibitem{molkenthin2020scaling}
N.~Molkenthin, M.~Schr{\"o}der, and M.~Timme.
\newblock Scaling laws of collective ride-sharing dynamics.
\newblock {\em Physical Review Letters}, 125(24):248302, 2020.

\bibitem{molkenthin2016scaling}
N.~Molkenthin and M.~Timme.
\newblock Scaling laws in spatial network formation.
\newblock {\em Physical review letters}, 117(16):168301, 2016.

\bibitem{muhle2023analytical}
S.~M{\"u}hle.
\newblock An analytical framework for modeling ride pooling efficiency and minimum fleet size.
\newblock {\em Multimodal Transportation}, 2(2):100080, 2023.

\bibitem{nicolaou2017chimera}
Z.~G. Nicolaou, H.~Riecke, and A.~E. Motter.
\newblock Chimera states in continuous media: Existence and distinctness.
\newblock {\em Physical review letters}, 119(24):244101, 2017.

\bibitem{SCHMAUS2025104726}
A.~Schmaus, F.~Creutzig, N.~Koch, F.~Nachtigall, and N.~Molkenthin.
\newblock An urban shared pooled mobility system cuts distance travelled by over 50
\newblock {\em Transportation Research Part D: Transport and Environment}, 144:104726, 2025.

\bibitem{schmaus_2026_20112639}
A.~Schmaus and N.~Molkenthin.
\newblock Recurrence analysis for ride-pooling systems, May 2026.

\bibitem{shi2021influence}
K.~Shi, R.~Shao, J.~De~Vos, L.~Cheng, and F.~Witlox.
\newblock The influence of ride-hailing on travel frequency and mode choice.
\newblock {\em Transportation Research Part D: Transport and Environment}, 101:103125, 2021.

\bibitem{tachet2017scaling}
R.~Tachet, O.~Sagarra, P.~Santi, G.~Resta, M.~Szell, S.~Strogatz, and C.~Ratti.
\newblock Scaling law of urban ride sharing.
\newblock {\em Scientific reports}, 7:42868, 2017.

\bibitem{zech2022collective}
R.~M. Zech, N.~Molkenthin, M.~Timme, and M.~Schr{\"o}der.
\newblock Collective dynamics of capacity-constrained ride-pooling fleets.
\newblock {\em Scientific reports}, 12(1):1--9, 2022.

\bibitem{zhang2021mechanism}
Y.~Zhang and A.~E. Motter.
\newblock Mechanism for strong chimeras.
\newblock {\em Physical review letters}, 126(9):094101, 2021.

\bibitem{zhong2026beyond}
L.~Zhong and L.~Cheng.
\newblock Beyond resilience curves: rethinking transport system resilience assessment, 2026.

\end{thebibliography}

\end{document}